


\font\mayusc=cmcsc10 


      \font \ninebf                 = cmbx9
      \font \ninei                  = cmmi9
      \font \nineit                 = cmti9
      \font \ninerm                 = cmr9
      \font \ninesans               = cmss10 at 9pt
      \font \ninesl                 = cmsl9
      \font \ninesy                 = cmsy9
      \font \ninett                 = cmtt9
      \font \fivesans               = cmss10 at 5pt
						\font \sevensans              = cmss10 at 7pt
      \font \sixbf                  = cmbx6
      \font \sixi                   = cmmi6
      \font \sixrm                  = cmr6
						\font \sixsans                = cmss10 at 6pt
      \font \sixsy                  = cmsy6
      \font \tams                   = cmmib10
      \font \tamss                  = cmmib10 scaled 700
						\font \tensans                = cmss10
    
      \skewchar\ninei='177 \skewchar\sixi='177
      \skewchar\ninesy='60 \skewchar\sixsy='60
      \hyphenchar\ninett=-1
      \def\newline{\hfil\break}%
      \catcode`@=11
      \def\folio{\ifnum\pageno<\z@
      \uppercase\expandafter{\romannumeral-\pageno}%
      \else\number\pageno \fi}
      \catcode`@=12 

      \newfam\sansfam
      \textfont\sansfam=\tensans\scriptfont\sansfam=\sevensans
      \scriptscriptfont\sansfam=\fivesans
      \def\sans{\fam\sansfam\tensans}


      \def\petit{\def\rm{\fam0\ninerm}%
      \textfont0=\ninerm \scriptfont0=
\sixrm \scriptscriptfont0=\fiverm
       \textfont1=\ninei \scriptfont1=
\sixi \scriptscriptfont1=\fivei
       \textfont2=\ninesy \scriptfont2=
\sixsy \scriptscriptfont2=\fivesy
       \def\it{\fam\itfam\nineit}%
       \textfont\itfam=\nineit
       \def\sl{\fam\slfam\ninesl}%
       \textfont\slfam=\ninesl
       \def\bf{\fam\bffam\ninebf}%
       \textfont\bffam=\ninebf \scriptfont\bffam=\sixbf
       \scriptscriptfont\bffam=\fivebf
       \def\sans{\fam\sansfam\ninesans}%
       \textfont\sansfam=\ninesans \scriptfont\sansfam=\sixsans
       \scriptscriptfont\sansfam=\fivesans
       \def\tt{\fam\ttfam\ninett}%
       \textfont\ttfam=\ninett
       \normalbaselineskip=11pt
       \setbox\strutbox=\hbox{\vrule height7pt depth2pt width0pt}%
       \normalbaselines\rm


      \def\bvec##1{{\textfont1=\tbms\scriptfont1=\tbmss
      \textfont0=\ninebf\scriptfont0=\sixbf
      \mathchoice{\hbox{$\displaystyle##1$}}{\hbox{$\textstyle##1$}}
      {\hbox{$\scriptstyle##1$}}{\hbox{$\scriptscriptstyle##1$}}}}}


.

					\mathchardef\gammav="0100
     \mathchardef\deltav="0101
     \mathchardef\thetav="0102
     \mathchardef\lambdav="0103
     \mathchardef\xiv="0104
     \mathchardef\piv="0105
     \mathchardef\sigmav="0106
     \mathchardef\upsilonv="0107
     \mathchardef\phiv="0108
     \mathchardef\psiv="0109
     \mathchardef\omegav="010A


					\mathchardef\Gammav="0100
     \mathchardef\Deltav="0101
     \mathchardef\Thetav="0102
     \mathchardef\Lambdav="0103
     \mathchardef\Xiv="0104
     \mathchardef\Piv="0105
     \mathchardef\Sigmav="0106
     \mathchardef\Upsilonv="0107
     \mathchardef\Phiv="0108
     \mathchardef\Psiv="0109
     \mathchardef\Omegav="010A



\font\grbfivefm=cmbx5
\font\grbsevenfm=cmbx7
\font\grbtenfm=cmbx10 
\newfam\grbfam
\textfont\grbfam=\grbtenfm
\scriptfont\grbfam=\grbsevenfm
\scriptscriptfont\grbfam=\grbfivefm

\font\calbfivefm=cmbsy10 at 5pt
\font\calbsevenfm=cmbsy10 at 7pt
\font\calbtenfm=cmbsy10 
\newfam\calbfam
\textfont\calbfam=\calbtenfm
\scriptfont\calbfam=\calbsevenfm
\scriptscriptfont\calbfam=\calbfivefm



      \def\bvec#1{{\textfont1=\tams\scriptfont1=\tamss
      \textfont0=\tenbf\scriptfont0=\sevenbf
      \mathchoice{\hbox{$\displaystyle#1$}}{\hbox{$\textstyle#1$}}
      {\hbox{$\scriptstyle#1$}}{\hbox{$\scriptscriptstyle#1$}}}}



\def\pmbf#1{\leavevmode\setbox0=\hbox{#1}%
\kern-.02em\copy0\kern-\wd0
\kern.04em\copy0\kern-\wd0
\kern-.02em\copy0\kern-\wd0
\kern-.03em\copy0\kern-\wd0
\kern.06em\box0 }



						\def\monthname{%
   			\ifcase\month
      \or Jan\or Feb\or Mar\or Apr\or May\or Jun%
      \or Jul\or Aug\or Sep\or Oct\or Nov\or Dec%
   			\fi
							}%
					\def\timestring{\begingroup
   		\count0 = \time
   		\divide\count0 by 60
   		\count2 = \count0   
   		\count4 = \time
   		\multiply\count0 by 60
   		\advance\count4 by -\count0   
   		\ifnum\count4<10
     \toks1 = {0}%
   		\else
     \toks1 = {}%
   		\fi
   		\ifnum\count2<12
      \toks0 = {a.m.}%
   		\else
      \toks0 = {p.m.}%
      \advance\count2 by -12
   		\fi
   		\ifnum\count2=0
      \count2 = 12
   		\fi
   		\number\count2:\the\toks1 \number\count4 \thinspace \the\toks0
					\endgroup}%

				\newskip\abovelistskip      \abovelistskip = .5\baselineskip 
				\newskip\interitemskip      \interitemskip = 0pt
				\newskip\belowlistskip      \belowlistskip = .5\baselineskip
				\newdimen\listleftindent    \listleftindent = 0pt
				\newdimen\listrightindent   \listrightindent = 0pt

				%


\def\petit{\def\rm{\fam0\ninerm}%
\textfont0=\ninerm \scriptfont0=\sixrm \scriptscriptfont0=\fiverm
\textfont1=\ninei \scriptfont1=\sixi \scriptscriptfont1=\fivei
\textfont2=\ninesy \scriptfont2=\sixsy \scriptscriptfont2=\fivesy
       \def\it{\fam\itfam\nineit}%
       \textfont\itfam=\nineit
       \def\sl{\fam\slfam\ninesl}%
       \textfont\slfam=\ninesl
       \def\bf{\fam\bffam\ninebf}%
       \textfont\bffam=\ninebf \scriptfont\bffam=\sixbf
       \scriptscriptfont\bffam=\fivebf
       \def\sans{\fam\sansfam\ninesans}%
       \textfont\sansfam=\ninesans \scriptfont\sansfam=\sixsans
       \scriptscriptfont\sansfam=\fivesans
       \def\tt{\fam\ttfam\ninett}%
       \textfont\ttfam=\ninett
       \normalbaselineskip=11pt
       \setbox\strutbox=\hbox{\vrule height7pt depth2pt width0pt}%
       \normalbaselines\rm
      \def\vec##1{{\textfont1=\tbms\scriptfont1=\tbmss
      \textfont0=\ninebf\scriptfont0=\sixbf
      \mathchoice{\hbox{$\displaystyle##1$}}{\hbox{$\textstyle##1$}}
      {\hbox{$\scriptstyle##1$}}{\hbox{$\scriptscriptstyle##1$}}}}}

      \def\footnoterule{\kern-3pt\hrule width 2cm\kern2.6pt}
      \newdimen\oldparindent\oldparindent=1.5em
      \parindent=1.5em
 
\newcount\footcount \footcount=0
\def\advftncnt{\advance\footcount by1\global\footcount=\footcount}
      \def\fnote#1{\advftncnt$^{\the\footcount}$\begingroup\petit
      \parfillskip=0pt plus 1fil
      \def\textindent##1{\hangindent0.5\oldparindent\noindent\hbox
      to0.5\oldparindent{##1\hss}\ignorespaces}%
 \vfootnote{$^{\the\footcount}$}
{#1\nullbox{0mm}{2mm}{0mm}\vskip-9.69pt}\endgroup}


      \def\item#1{\par\noindent
      \hangindent6.5 mm\hangafter=0
      \llap{#1\enspace}\ignorespaces}
      
      \def\leaderfill{\kern0.5em\leaders
\hbox to 0.5em{\hss.\hss}\hfill\kern
      0.5em}
						\def\hb{\hfill\break}

    \def\centerrule#1{\centerline{\kern#1\hrulefill\kern#1}}


      \def\boxit#1{\vbox{\hrule\hbox{\vrule\kern3pt
						\vbox{\kern3pt#1\kern3pt}\kern3pt\vrule}\hrule}}

      \def\tightboxit#1{\vbox{\hrule\hbox{\vrule
						\vbox{#1}\vrule}\hrule}}

      \def\looseboxit#1{\vbox{\hrule\hbox{\vrule\kern5pt
						\vbox{\kern5pt#1\kern5pt}\kern5pt\vrule}\hrule}}

      \def\youboxit#1#2{\vbox{\hrule\hbox{\vrule\kern#2
						\vbox{\kern#2#1\kern#2}\kern#2\vrule}\hrule}}



			\def\whitetile#1#2#3{\setbox0=\null
			\ht0=#1 \dp0=#2\wd0=#3 \setbox1=
\hbox{\tightboxit{\box0}}\lower#2\box1}

			\def\nullbox#1#2#3{\setbox0=\null
			\ht0=#1 \dp0=#2\wd0=#3\box0}




\def\fig{\leavevmode Fig.}
\def\figs{\leavevmode Figs.}
\def\equ{\leavevmode Eq.}

\def\equs{\leavevmode Eqs.}
\def\sect{\leavevmode Sect.}
\def\subsect{\leavevmode Subsect.}

\def\equn#1{\ifmmode \eqno{\rm #1}\else \equ~#1\fi}



\def\tev{\ifmmode \mathop{\rm TeV}\nolimits\else {\rm TeV}\fi}
\def\gev{\ifmmode \mathop{\rm GeV}\nolimits\else {\rm GeV}\fi}
\def\mev{\ifmmode \mathop{\rm MeV}\nolimits\else {\rm MeV}\fi}
\def\kev{\ifmmode \mathop{\rm keV}\nolimits\else {\rm keV}\fi}
\def\ev{\ifmmode \mathop{\rm eV}\nolimits\else {\rm eV}\fi}

\def\chidof{\ifmmode
\mathop\chi^2/{\rm d.o.f.}\else $\chi^2/{\rm d.o.f.}\null$\fi}

\def\msbar{\ifmmode
\mathop{\overline{\rm MS}}\else$\overline{\rm MS}$\null\fi}


\def\physmatex{P\kern-.14em\lower.5ex\hbox{\sevenrm H}ys
\kern -.35em \raise .6ex \hbox{{\sevenrm M}a}\kern -.15em
 T\kern-.1667em\lower.5ex\hbox{E}\kern-.125emX\null}%

\def\ref#1{$^{[#1]}$\relax}

\def\ajnyp#1#2#3#4#5{
\frenchspacing{\mayusc #1}, {\sl#2}, {\bf #3} ({#4}) {#5}}













\def\ddal{\mathop{\vrule height 7pt depth0.2pt
\hbox{\vrule height 0.5pt depth0.2pt width 6.2pt}
\vrule height 7pt depth0.2pt width0.8pt
\kern-7.4pt\hbox{\vrule height 7pt depth-6.7pt width 7.pt}}}
\def\sdal{\mathop{\kern0.1pt\vrule height 4.9pt depth0.15pt
\hbox{\vrule height 0.3pt depth0.15pt width 4.6pt}
\vrule height 4.9pt depth0.15pt width0.7pt
\kern-5.7pt\hbox{\vrule height 4.9pt depth-4.7pt width 5.3pt}}}
\def\ssdal{\mathop{\kern0.1pt\vrule height 3.8pt depth0.1pt width0.2pt
\hbox{\vrule height 0.3pt depth0.1pt width 3.6pt}
\vrule height 3.8pt depth0.1pt width0.5pt
\kern-4.4pt\hbox{\vrule height 4pt depth-3.9pt width 4.2pt}}}




\mathchardef\lap='0001


\def\lsim{\mathop{\setbox0=\hbox{$\displaystyle 
\raise2.2pt\hbox{$\;<$}\kern-7.7pt\lower2.6pt\hbox{$\sim$}\;$}
\box0}}
\def\gsim{\mathop{\setbox0=\hbox{$\displaystyle 
\raise2.2pt\hbox{$\;>$}\kern-7.7pt\lower2.6pt\hbox{$\sim$}\;$}
\box0}}

\def\gsimsub#1{\mathord{\vtop to0pt{\ialign{##\crcr
$\hfil{{\mathop{\setbox0=\hbox{$\displaystyle 
\raise2.2pt\hbox{$\;>$}\kern-7.7pt\lower2.6pt\hbox{$\sim$}\;$}
\box0}}}\hfil$\crcr\noalign{\kern1.5pt\nointerlineskip}
$\hfil\scriptstyle{#1}{}\kern1.5pt\hfil$\crcr}\vss}}}

\def\lsimsub#1{\mathord{\vtop to0pt{\ialign{##\crcr
$\hfil\displaystyle{\mathop{\setbox0=\hbox{$\displaystyle 
\raise2.2pt\hbox{$\;<$}\kern-7.7pt\lower2.6pt\hbox{$\sim$}\;$}
\box0}}
\def\gsim{\mathop{\setbox0=\hbox{$\displaystyle 
\raise2.2pt\hbox{$\;>$}\kern-7.7pt\lower2.6pt\hbox{$\sim$}\;$}
\box0}}\hfil$\crcr\noalign{\kern1.5pt\nointerlineskip}
$\hfil\scriptstyle{#1}{}\kern1.5pt\hfil$\crcr}\vss}}}

\def\dd{{\rm d}}





\def\frac#1#2{{#1\over#2}}
\def\dfrac#1#2{{\displaystyle{#1\over#2}}}
\def\tfrac#1#2{{\textstyle{#1\over#2}}}
\def\ffrac#1#2{\leavevmode
   \kern.1em \raise .5ex \hbox{\the\scriptfont0 #1}%
   \kern-.1em $/$%
   \kern-.15em \lower .25ex \hbox{\the\scriptfont0 #2}%
}%



\def\brochureb#1#2#3{\pageno#3
\headline={\ifodd\pageno{\rheadline}
\else\lheadline\fi}
\def\rheadline{\hfil -{#2}-\hfil}
\def\lheadline{\hfil-{#1}-\hfil}
\footline={\hss -- \number\pageno\ --\hss}
\voffset=2\baselineskip}

\def\nada{\phantom{M}\kern-1em}
\def\brochureendcover#1{\vfill\eject\pageno=1{\nada#1}\vfill\eject}





\def\chapterb#1#2#3{\pageno#3
\headline={\ifodd\pageno{\ifnum\pageno=#3\hfil\else\rheadline\fi}
\else\lheadline\fi}
\def\rheadline{\hfil -{#2}-\hfil}
\def\lheadline{\hfil-{#1}-\hfil}
\footline={\hss -- \number\pageno\ --\hss}
\voffset=2\baselineskip}


\def\bookendchapter{\ifodd\pageno
 \vfill\eject\footline={\hfill}\headline={\hfill}\null \vfill\eject
 \else\vfill\eject \fi}

\def\obookendchapter{\ifodd\pageno\vfill\eject
 \else\vfill\eject\null \vfill\eject\fi}


\def\booksection#1{
\setbox0=\vbox{\hsize=0.85\hsize\tolerance=500\raggedright\hfuzz=6mm
\noindent{\medfib #1}\medskip}\goodbreak\vskip0.6cm\box0
\nobreak
\noindent}
\def\booksubsection#1{
\setbox0=\vbox{\hsize=0.85\hsize\tolerance=400\raggedright\hfuzz=4mm
\noindent{\fib #1}\smallskip}\goodbreak\vskip0.45cm\box0
\nobreak
\noindent}




\def\figurasc#1#2{\petit{\noindent\sc#1}\ #2}

\def\captiontype{\tolerance=800\hfuzz=1mm\raggedright\noindent}



\def\abstracttype#1{
\hsize0.7\hsize\tolerance=800\hfuzz=0.5mm \noindent{\fib #1}\par
\medskip\petit}


\def\hb{\hfill\break}


\font\twelverm=cmr12 
\font\smallsc=cmcsc10 at 9pt 
\font\fib=cmfib8
\font\medfib=cmfib8 at 9pt


\font\sc=cmcsc10 

\font\addressfont=cmbxti10 at 9pt


\catcode`@=11 

\newdimen\pagewidth \newdimen\pageheight \newdimen\ruleht
 \maxdepth=2.2pt  \parindent=3pc
\pagewidth=\hsize \pageheight=\vsize \ruleht=.4pt
\abovedisplayskip=6pt plus 3pt minus 1pt
\belowdisplayskip=6pt plus 3pt minus 1pt
\abovedisplayshortskip=0pt plus 3pt
\belowdisplayshortskip=4pt plus 3pt

\newinsert\margin
\dimen\margin=\maxdimen




\newdimen\paperheight \paperheight = \vsize
\def\topmargin{\afterassignment\@finishtopmargin \dimen0}%
\def\@finishtopmargin{%
  \dimen2 = \voffset		
  \voffset = \dimen0 \advance\voffset by -1in
  \advance\dimen2 by -\voffset	
  \advance\vsize by \dimen2	
}%
\def\advancetopmargin{%
  \dimen0 = 0pt \afterassignment\@finishadvancetopmargin \advance\dimen0
}%
\def\@finishadvancetopmargin{%
  \advance\voffset by \dimen0
  \advance\vsize by -\dimen0
}%
\def\bottommargin{\afterassignment\@finishbottommargin \dimen0}%
\def\@finishbottommargin{%
  \@computebottommargin		
  \advance\dimen2 by -\dimen0	
  \advance\vsize by \dimen2	
}%
\def\advancebottommargin{%
  \dimen0 = 0pt\afterassignment\@finishadvancebottommargin \advance\dimen0
}%
\def\@finishadvancebottommargin{%
  \advance\vsize by -\dimen0
}%
\def\@computebottommargin{%
  \dimen2 = \paperheight	
  \advance\dimen2 by -\vsize	
  \advance\dimen2 by -\voffset	
  \advance\dimen2 by -1in	
}%
\newdimen\paperwidth \paperwidth = \hsize
\def\leftmargin{\afterassignment\@finishleftmargin \dimen0}%
\def\@finishleftmargin{%
  \dimen2 = \hoffset		
  \hoffset = \dimen0 \advance\hoffset by -1in
  \advance\dimen2 by -\hoffset	
  \advance\hsize by \dimen2	
}%
\def\advanceleftmargin{%
  \dimen0 = 0pt \afterassignment\@finishadvanceleftmargin \advance\dimen0
}%
\def\@finishadvanceleftmargin{%
  \advance\hoffset by \dimen0
  \advance\hsize by -\dimen0
}%
\def\rightmargin{\afterassignment\@finishrightmargin \dimen0}%
\def\@finishrightmargin{%
  \@computerightmargin		
  \advance\dimen2 by -\dimen0	
  \advance\hsize by \dimen2	
}%
\def\advancerightmargin{%
  \dimen0 = 0pt \afterassignment\@finishadvancerightmargin \advance\dimen0
}%
\def\@finishadvancerightmargin{%
  \advance\hsize by -\dimen0
}%
\def\@computerightmargin{%
  \dimen2 = \paperwidth		
  \advance\dimen2 by -\hsize	
  \advance\dimen2 by -\hoffset	
  \advance\dimen2 by -1in	
}%

\def\onepageout#1{\shipout\vbox{ 
    \offinterlineskip 
    \vbox to 3pc{ 
      \iftitle 
        \global\titlefalse 
        \setcornerrules 
      \else\ifodd\pageno \rightheadline\else\leftheadline\fi\fi
      \vfill} 
    \vbox to \pageheight{
      \ifvoid\margin\else 
        \rlap{\kern31pc\vbox to\z@{\kern4pt\box\margin \vss}}\fi
      #1 
      \ifvoid\footins\else 
        \vskip\skip\footins \kern-3pt
        \hrule height\ruleht width\pagewidth \kern-\ruleht \kern3pt
        \unvbox\footins\fi
      \boxmaxdepth=\maxdepth
      } 
    }
  \advancepageno}

\def\setcornerrules{\hbox to \pagewidth{\vrule width 1pc height\ruleht
    \hfil \vrule width 1pc}
  \hbox to \pagewidth{\llap{\sevenrm(page \folio)\kern1pc}%
    \vrule height1pc width\ruleht depth\z@
    \hfil \vrule width\ruleht depth\z@}}
\newbox\partialpage






\input epsf.sty
\raggedbottom
\footline={\hfill}
\rightline{FTUAM 99-8}
\rightline{UG-FT-97/99}
\rightline{hep-ph/9904344}
\rightline{April 14, 1999}
\rightline{(Revised August, 11, 1999)}
\bigskip
\hrule height .3mm
\vskip.6cm
\centerline{{\twelverm Calculation of 
Electroproduction to NNLO and Precision
 Determination of  $\alpha_s$}}
\medskip
\centerrule{.7cm}
\vskip1cm
\setbox8=\vbox{\hsize65mm {\noindent\fib J. Santiago} 
\vskip .1cm
\noindent{\addressfont Departamento de F\'\i sica Te\'orica\hb y del Cosmos,\hb
Universidad de Granada,\hb
E-18071, Granada, Spain.}}
\centerline{\box8}
\medskip
\setbox7=\vbox{\hsize65mm \fib and} 
\centerline{\box7}
\medskip
\setbox9=\vbox{\hsize65mm {\noindent\fib F. J. 
Yndur\'ain} 
\vskip .1cm
\noindent{\addressfont Departamento de F\'{\i}sica Te\'orica, C-XI,\hb
 Universidad Aut\'onoma de Madrid,\hb
 Canto Blanco,\hb
E-28049, Madrid, Spain.}\hb}
\smallskip
\centerline{\box9}
\bigskip
\setbox0=\vbox{\abstracttype{Abstract}We use the known values of the
 two loop Wilson coefficients 
and the three loop anomalous dimension matrix $\gamma(n)$ to perform 
a next-to-next-to leading order (NNLO) calculation of $ep$ deep 
inelastic scattering. Because  $\gamma(n)$ is only known for a few
values of $n$,  
the method of average reconstruction has to be used, which 
leaves 102 effective experimental points, for 12 parameters: 
the QCD mass $\Lambdav$, and 11 initial values
 for the moments of the structure functions. 
The data points spread in the range of momenta
 $2.5\;{\gev}^2\leq Q^2\leq 230\;{\gev}^2$.
 
The \chidof\ decreases substantially when going from LO to NLO, and also from NLO to 
NNLO (although only a little now) to $\chidof=79.2/(102-12)$.
 The favoured value of $\Lambdav$ is
$$\Lambdav(n_f=4,3\;{\rm loop})=282.7\pm35.1\;\mev,$$
corresponding to the value of the coupling at the Z mass of
$$\alpha^{(\rm 3\;loop)}_s(M_Z^2)=0.1172\pm0.0024. $$
The calculation, which constitutes a very  
precise test of QCD, includes target mass corrections; the error takes
into account  
 experimental errors and 
 higher twist effects  among other estimated 
theoretical errors.}
\centerline{\box0}
\brochureendcover{Typeset with \physmatex}
\pageno=1
\brochureb{\smallsc j. santiago and f. j.  yndur\'ain}{\smallsc calculation of 
electroproduction to nnlo and precision determination of $\alpha_s$}{1}

\booksection{1 Introduction}
Deep inelastic scattering (DIS), in particular of electrons (or muons) 
on protons, constituted one 
of the first probes of hadron structure. The calculation of QCD-induced scaling 
violations in the structure functions\ref{1} yielded some of the 
earliest qualitative checks of 
the quark-gluon theory of hadron interactions, as well as providing the first two loop  
determinations of the strong coupling constant\ref{2}.
 In this  last ref.~2, the two loop 
results for the nonsinglet anomalous dimensions\ref{2,3} were used,
 together with the one loop Wilson coefficients\ref{4} to get a calculation
 of the nonsinglet  structure functions  to next to leading order (NLO).\fnote{To 
leading order (LO) one requires the one loop anomalous dimension and the tree level 
Wilson coefficients. Likewise, a NLO calculation
 uses one loop Wilson coefficients and two 
loop anomalous dimension and the NNLO one (which is the subject of the 
present paper) implies two loop Wilson coefficients and 
three loop anomalous dimensions.} Because
 both anomalous dimensions and coefficient functions 
were known completely, it was possible to use a method 
devised by Gross\ref{5} to LO, generalized in ref.~2 to NLO, to obtain 
an exact, pointlike (in $x$) reconstruction of the nonsinglet part of $F_2(x,Q^2)$, 
and of $xF_3(x,Q^2)$.

Since then the calculations have been extended to include
 singlet anomalous dimensions (ref.~6; see ref.~7
 for a collection of two loop formulas). 
Furthermore,  the  two loop Wilson coefficients are also known\ref{8} 
so that, to perform a NNLO calculation, only the three loop
 anomalous dimensions are needed. 
There does not exist at present a full calculation of these;
 but a partial computation 
has been made available recently.\ref{9} 
This has been used in some  papers (see e.g. ref.~10 and work quoted there) 
to perform an evaluation of deep inelastic structure functions to 
next-to-next-to-leading order, NNLO.
 These evaluations, however, 
present a number of shortcomings. First of all, the values of
 the moments which are not known are found 
by interpolation of the known 
moments; and likewise the {\sl experimental} 
values of the moments, with which to compare the theory, are obtained also 
interpolating and extrapolating experimental data.
 These procedures are dangerous in 
that they imply unknown systematic errors,
 thus putting in jeopardy an already delicate 
calculation (NNLO effects 
are in themselves small, generally speaking).
 Perhaps more important, only {\sl nonsinglet} 
structure functions (specifically, 
$xF_3$ for $\nu$ scattering in ref.~10) were
 considered: while the best data exist for 
the $ep$ (or $\mu p$) structure function $F_2$, 
with a strong singlet component.

In the present paper we improve on this in the following ways.
 First of all, we 
extend the calculation to include  singlet as well as nonsinglet structure 
functions. This allows us to use both the 
very precise, comparatively low energy, electroproduction SLAC data, 
and the recent, very high energy HERA results, plus 
some intermediate energy muon-production data. This provides a huge range of 
$Q^2$ values. Secondly, we 
employ the method of Bernstein polynomials.
 This method was developed in ref.~11 
for calculating {\sl averages} of the structure functions 
around values of $x$ where 
experimental data are available, in terms of only a finite set 
of known moments: thus at the same time 
avoiding the problem of interpolating the theoretical values of the  
anomalous dimension, and greatly diminishing
 the errors inherent to the calculation 
of the experimental input, as only known sets
 of data points are essentially relevant.

In this way we find a  precise comparison of electroproduction 
DIS data with theory, 
obtaining in particular a very accurate evaluation of the QCD coupling and mass.
 With respect 
to the last we find
$$\eqalign{\alpha_s^{(\rm 3\;loop)}(M_Z^2)=&
0.1172\pm0.0017\;(\hbox{statistical})\pm0.0017\;(\hbox{systematic})\cr
=&0.1172\pm0.0024,\cr}$$
to which corresponds the value and error of the mass parameter
$$\Lambdav(n_f=4,3\,{\rm loop})=283\pm35\;\mev.$$

\booksection{2 The evolution equations}
 To describe the calculation, let us 
establish some notation first. We split $F_2$ 
into a singlet and a nonsinglet part,
$$F_2(x,Q^2)=F_S(x,Q^2)+F_{NS}(x,Q^2).\equn{(2.1)}$$
We also define the {\sl gluon 
structure function}, $F_G(x,Q^2)\equiv xG(x,Q^2)$, $G$ being the 
gluon density. Finally, we form the vector with components $F_S,\,F_G$:
$$F(x,Q^2)=\pmatrix{F_S(x,Q^2)\cr F_G(x,Q^2)\cr}.$$ 
We obtain the {\sl moments} by projecting these with powers of $x$:
$$\eqalign{\mu_{NS}(n;Q^2)=&\int_0^1\dd x\,x^{n-2}F_{NS}(x,Q^2);\cr
\mu_i(n;Q^2)=&\int_0^1\dd x\,x^{n-2}F_i(x,Q^2);\quad i=S,\,G.\cr}\equn{(2.2)}$$
If we write a light cone expansion for the moments, 
this is given in terms of the Wilson coefficients and the anomalous dimensions of 
appropriate sets of operators (for details on this,
 see any text on QCD, e.g., ref.~12). 
Let $t=\tfrac{1}{2}\log(Q^2/\nu^2)$, with $\nu$ 
the renormalization scale, and 
$a=\alpha_s(Q^2)/4\pi,\;a'=\alpha_s({Q'}^2)/4\pi$.
 For the nonsinglet component, we have
$$\mu_{NS}(n;Q^2)=C_{NS}\left(n,\alpha_s(Q^2)/4\pi\right)
\exp\left[-\int_0^t\dd t'\,\gamma_{NS}(a')\right]A_{NS}(\nu^2)
\equn{(2.3a)}$$
whereas the singlet equations are of a matrix character,
$$\mu_i(n;Q^2)=\sum_j C_j\left(n,\alpha_s(Q^2)/4\pi\right)
\left\{T\exp\left[-\int_0^t\dd t'\,{\gamma}(a')\right]{ A}(\nu^2)\right\}_{ji}
\equn{(2.3b)}$$
The $\gamma_{NS},\,{\gamma}$ are the NS, S
 anomalous dimension and anomalous dimension matrix, and 
the $A_{NS},\,A$ unknown matrix elements of certain operators. What 
the calculations of Larin et al.\ref{9} provide is the expression of 
the gammas to three
 loops (third order in $\alpha_s$) for $n=2,\,4,\,6,\,8$.\fnote{Note that 
the conventions of ref.~9 differ from ours here in that 
their $\gamma$'s are a factor of two larger than ours:
$$2\gamma^{\rm ref.\,9}(n)=\gamma^{\rm here}(n).$$}

Before discussing how to use this information, let us write 
 the equations that follow from 
\equs~(2.3). We will consider in some detail only the singlet component; the NS one 
is fairly trivial. We start by completing
 a two by two matrix $ C$ from $C_S,\,C_G$,
$$C=\pmatrix{C_S&C_{12}\cr C_G&C_{22}\cr}\equiv
\pmatrix{C_{11}&C_{12}\cr C_{21}&C_{22}\cr}$$
by requiring it to commute with the anomalous dimension matrix:
$$[C,\gamma\,]=0$$
(the index $n$ will be omitted in some of the formulas to lighten the notation). 
We define 
$$D(a)=\dfrac{\gamma(a)}{2\beta(a)}\equn{(2.4)}$$
and write the series expansions,
$$\eqalign{C(a)=&1+C^{(1)}\,a+C^{(2)}\,a^2+\dots\cr
\gamma(a)=&\gamma^{(0)}\,a+\gamma^{(1)}\,a^2+ \gamma^{(2)}\,a^3+\dots\cr
-\beta(a)=&\beta_0a^2+\beta_1a^3+\beta_2a^4+\dots,\cr
D(a)=&\dfrac{1}{a}D^{(0)}+D^{(1)}+D^{(2)}a+\dots;}
\equn{(2.5)}$$
$$\eqalign{D^{(0)}=\dfrac{-1}{2\beta_0}\gamma^{(0)};\;
D^{(1)}=\dfrac{-1}{2\beta_0}\left(\gamma^{(1)}-
\dfrac{\beta_1}{\beta_0}\gamma^{(0)}\right);\cr
D^{(2)}=\dfrac{-1}{2\beta_0}
\left[\gamma^{(2)}-\dfrac{\beta_1}{\beta_0}\gamma^{(1)}+
\left(\dfrac{\beta_1^2}{\beta_0}-
\dfrac{\beta_2}{\beta_0}\right)\gamma^{(0)}\right];\cdots\,.}
\equn{(2.6)}$$ 
The commutativity of $ C$ and $\gamma$, order by order in 
perturbation theory, implies the equations
$$\eqalign{C^{(1)}_{21}
=&\dfrac{\gamma^{(0)}_{21}}{\gamma^{(0)}_{12}}\,C^{(1)}_{12},\cr
C^{(1)}_{22}=&C^{(1)}_{11}+
\dfrac{\gamma^{(0)}_{22}-\gamma^{(0)}_{11}}{\gamma^{(0)}_{12}}\,C^{(1)}_{12};\cr
C^{(2)}_{21}=&\dfrac{C^{(1)}_{12}\gamma^{(1)}_{21}+
C^{(2)}_{12}\gamma^{(0)}_{21}-
C^{(1)}_{21}\gamma^{(1)}_{12}}{\gamma^{(0)}_{12}},\cr
C^{(2)}_{22}=&C^{(2)}_{11}+
\dfrac{\gamma^{(0)}_{22}-\gamma^{(0)}_{11}}{\gamma^{(0)}_{12}}\,C^{(2)}_{12}+
\dfrac{\gamma^{(1)}_{22}-\gamma^{(1)}_{11}}{\gamma^{(0)}_{12}}\,C^{(1)}_{12}+
\dfrac{C^{(1)}_{11}-C^{(1)}_{22}}{\gamma^{(0)}_{12}}\gamma^{(1)}_{12}.\cr
}
\equn{(2.7)}$$
These equations allow us to find the $C_{21},\,C_{22}$ in terms of the
 $C_{11},\,C_{12}$ and the 
$\gamma_{ij}$ which in turn are taken from ref.~9.

For the moments equations, we define the 
matrix $ S$ that diagonalizes the LO anomalous dimension matrix, so that
$$S^{-1}D^{(0)}S=\hat{D}^{(0)}=
\pmatrix{d_{+}&0\cr0&d_-},\;d_+>d_-.\equn{(2.8a)}$$
 We take it to be
$$S=\pmatrix{1&\dfrac{D^{(0)}_{12}}{d_--d_+}\cr
\dfrac{d_+-D^{(0)}_{11}}{D^{(0)}_{12}}&\dfrac{d_--D^{(0)}_{11}}{d_--d_+}},\;
S_{11}=\det S=1,\equn{(2.8b)}$$
and we define also
$$\eqalign{{ S}^{-1}D^{(N)}{ S}\equiv\bar{D}^{(N)},\;
{ S}^{-1}M^{(N)}{ S}\equiv\bar{M}^{(N)},\;{ S}^{-1}\gamma^{(N)}{ S}
\equiv\bar\gamma^{(N)}\quad {\rm etc.}\cr
(\bar{D}^{(0)}=\hat{D}^{(0)}).}\equn{(2.8c)}$$
Here the matrix $M$ is defined to be such that
$$\dfrac{\partial}{\partial a}\left\{a^{D^{(0)}}\,M(a)C^{-1}(a)\mu(a)\right\}=0,
\equn{(2.9)}$$
and it has the series expansion
$$M(a)=1+M^{(1)}a+M^{(2)}a^2+\cdots.$$
From \equs~(2.8c),~(2.9) we find the explicit values of the $M$:
$$\bar{M}^{(1)}=\pmatrix{\bar{D}^{(1)}_{11}&
\dfrac{1}{1+d_+-d_-}\bar{D}^{(1)}_{12}\cr
\dfrac{1}{1+d_--d_+}\bar{D}^{(1)}_{21}&\bar{D}^{(1)}_{22}\cr},
\equn{(2.10a)}$$
$$\bar{M}^{(2)}=\pmatrix{\tfrac{1}{2}\left[\bar{D}^{(2)}_{11}+
(\bar{M}^{(1)}\bar{D}^{(1)})_{11}\right]
&\dfrac{1}{2+d_+-d_-}\,\left[\bar{D}^{(2)}_{12}+
(\bar{M}^{(1)}\bar{D}^{(1)})_{12}\right]\cr 
\dfrac{1}{2+d_--d_+}\,\left[\bar{D}^{(2)}_{21}+
(\bar{M}^{(1)}\bar{D}^{(1)})_{21}\right]
&\tfrac{1}{2}\left[\bar{D}^{(2)}_{22}+(\bar{M}^{(1)}\bar{D}^{(1)})_{22}\right]\cr}.
\equn{(2.10b)}$$

The evolution equation may then be written as, to NNLO,
$$\eqalign{\mu(n,a)=&C(n,a)S(n)\bar{M}(n,a)^{-1}
\left(a_0/a\right)^{\hat{D}^{(0)}(n)}
\bar{M}(n,a_0)S^{-1}(n)C(n,a_0)^{-1}\mu(n,a_0),\cr
a=&\alpha_s(Q^2)/4\pi,\quad a_0=\alpha_s(Q^2_0)/4\pi\cr}
\equn{(2.11)}$$
and $\alpha_s$ is to be calculated to three loops:
$$\alpha_s(Q^2)=\dfrac{4\pi}{\beta_0L}\left\{1-\dfrac{\beta_1\log L}{\beta_0^2L}+
\dfrac{\beta_1^2\log^2L-\beta_1^2\log L+\beta_2\beta_0-\beta_1^2}{\beta_0^4L^2}\right\}
$$
with
$$\eqalign{L=&\log\dfrac{Q^2}{\Lambdav^2};\quad
\beta_0=11-\tfrac{2}{3}n_f,\cr
\beta_1=&102-\tfrac{38}{3}n_f,\quad
\beta_2=\tfrac{2857}{2}-\tfrac{5033}{18}n_f+\tfrac{325}{54}n_f^2.\cr}$$
We work in the \msbar\ scheme throughout this paper.
For the NS component the corresponding equation is
$$\eqalign{\mu_{NS}(n,Q^2)=&
\left(\dfrac{\alpha_s(Q_0^2)}{\alpha_s(Q^2)}
\right)^{-\gamma_{NS}^{(0)}(n)/2\beta_0}\cr
\times&\dfrac{1+B_n^{(1)}\alpha_s(Q^2)/4\pi+B_n^{(2)}(\alpha_s(Q^2)/4\pi)^2}{1
+B_n^{(1)}\alpha_s(Q^2_0)/4\pi+B_n^{(2)}
(\alpha_s(Q^2_0)/4\pi)^2}\mu_{NS}(n,Q^2_0).\cr}
\equn{(2.12a)}$$
The explicit values of the $B_n^{(i)}$, expressible in 
terms of the $\gamma_{NS}^{(i)},\,C_{NS}^{(i)}(n)$, may 
be found in ref.~10. We can rewrite (2.12a) expanding the denominator 
as
$$\eqalign{\mu_{NS}(n,Q^2)=&
\left(\dfrac{\alpha_s(Q_0^2)}{\alpha_s(Q^2)}
\right)^{-\gamma_{NS}^{(0)}(n)/2\beta_0}\cr
\times&\Big\{1+B_n^{(1)}\alpha_s(Q^2)/4\pi-B_n^{(1)}\alpha_s(Q^2_0)/4\pi\cr
+&B_n^{(2)}\left(\alpha_s(Q^2)/4\pi\right)^2-B_n^{(2)}
\left(\alpha_s(Q^2_0)/4\pi\right)^2
+\left(B_n^{(1)}\alpha_s(Q^2_0)/4\pi\right)^2\cr
-&
\left(B_n^{(1)}\alpha_s(Q^2_0)/4\pi\right)
\left(B_n^{(1)}\alpha_s(Q^2)/4\pi)\right)\Big\}\mu_{NS}(n,Q^2_0).\cr}
\equn{(2.12b)}$$
The difference between (2.12b) and (2.12a) is of order $\alpha_s^3$, and may be used 
to estimate the effect of higher order ({\sl N}NNLO) corrections.

A last preliminary point to be discussed is that of the number of flavours. 
We will be working with the moments $\mu(n,Q^2)$, so we 
have only one momentum variable. We then split the $Q^2$
 range into the following two intervals:
$$Q^2\lsim m_b^2\;({\rm I});\quad m_b^2\lsim Q^2\;{(\rm II)}.$$
The values of $Q^2$ we will be using 
will be much less 
than $m_t^2$, and larger  than or of the order of $m_c^2$ so 
we need not consider $c,\,t$ quark thresholds. 
Then, in region (I) we take $n_f=4$ and in region 
(II), $n_f=5$. The matching will be carried 
over following the prescription of ref.~13: to NNLO,
$$\beta_0^{n_f+1}\log\dfrac{\Lambdav^2(n_f+1)}{\Lambdav^2(n_f)}=
(\beta_0^{n_f+1}-\beta_0^{n_f})L_h+\delta_{NLO}+\delta_{NNLO}
\equn{(2.13a)}$$
where
$$\eqalign{\delta_{NLO}=&
(b_1^{n_f+1}- b_1^{n_f})\log L_h-b_1^{n_f+1}
\log\dfrac{\beta_0^{n_f+1}}{\beta_0^{n_f}},\cr
\delta_{NNLO}=&\dfrac{1}{\beta_0^{n_f}L_h}
\left[(b_1^{n_f+1}- b_1^{n_f})b_1^{n_f}\log L_h+
(b_1^{n_f+1})^2-(b_1^{n_f})^2+b_2^{n_f}-b_2^{n_f+1}+\tfrac{7}{24}\right].\cr}
\equn{(2.13b)}$$
Here,
$$L_h=\log\left[m^2(n_f+1)/\Lambdav^2(n_f)\right],\quad b_i=\beta_i/\beta_0$$
and $m(n_f+1)$ is the pole mass of the $(n_f+1)$th quark.
\booksection{3. Experimental input}
\booksubsection{3.1 The method of Bernstein polynomials}
Because, for a given value of $Q^2$, only a limited 
number of experimental points, covering a partial range of values of $x$, 
are available, one cannot simply use the moments equations.
 A method devised to deal with a situation like the present one is that 
of averages with the (modified) Bernstein
 polynomials.\fnote{We call the polynomials 
modified because, since only even moments are known, we 
have to consider polynomials in the variable $x^2$. For more 
details on the method, see refs.~11,~14.} We define the (Bernstein) polynomials as
$$\eqalign{p_{nk}(x)=&
\dfrac{2\Gammav(n+\tfrac{3}{2})}
{\Gammav(k+\tfrac{1}{2})\Gammav(n-k+1)}x^{2k}(1-x^2)^{n-k}\cr
=&\dfrac{2(n-k)!\Gammav(n+\tfrac{3}{2})}{\Gammav(k+\tfrac{1}{2})\Gammav(n-k+1)}
\sum_{l=0}^{n-k}\dfrac{(-1)^l}{l!(n-k-l)!}x^{2(k+l)};\quad k\leq n.\cr}
\equn{(3.1)}$$
These polynomials have a number of useful properties. First, they are positive and
 have a single maximum located at 
$$\bar{x}_{nk}=\dfrac{\Gammav(k+1)\Gammav(n+\tfrac{3}{2})}
{\Gammav(k+\tfrac{1}{2})\Gammav(n+2)};$$
 they are concentrated around this point,
 with a spread of 
$$\lap x_{nk}=
\sqrt{\dfrac{k+\tfrac{1}{2}}{n+\tfrac{3}{2}}-
\left[\dfrac{\Gammav(k+1)\Gammav(n+\tfrac{3}{2})}
{\Gammav(k+\tfrac{1}{2})\Gammav(n+2)}\right]^2},$$
 and they are normalized to unity, 
$\int^1_0\dd x\,p_{nk}(x)=1$. Therefore, the integral
$$\int_0^1\dd x\,p_{nk}(x)\varphi(x)$$
represents an average of the 
function $\varphi(x)$ in 
the region $\bar{x}_{nk}-\tfrac{1}{2}\lap x_{nk}\lsim
 x\lsim\bar{x}_{nk}+\tfrac{1}{2}\lap x_{nk}$;
 the 
values of the function $\varphi(x)$ outside this interval contribute  little
 to the integral, as $p_{nk}(x)$ 
decreases to zero very quickly there. Finally, 
and using the binomial expansion in \equn{(3.1)}, it follows that 
the averages with the $p_{nk}$ of a function can be obtained in terms of 
its {\sl even} moments:
$$\eqalign{\int_0^1\dd x\,p_{nk}(x)\varphi(x)=&
\dfrac{2(n-k)!\Gammav(n+\tfrac{3}{2})}{\Gammav(k+\tfrac{1}{2})\Gammav(n-k+1)}
\sum_{l=0}^{n-k}\dfrac{(-1)^l}{l!(n-k-l)!}\,\varphi_{2k+2l},\cr
\varphi_{2l}=&\int_0^1\dd x\,x^{2l}\varphi(x).\cr}$$

 We will thus consider our 
experimental input to be given by averages
$$F^{\rm(exp)}_{nk}(Q^2)\equiv
\int_0^1\dd x\,p_{nk}(x)F_2^{\rm(exp)}(x,Q^2),
\equn{(3.2)}$$
with $F_2^{\rm(exp)}(x,Q^2)$ the experimental structure function.

\booksubsection{3.2 Calculation of the experimental averages} 
Even with the method of Bernstein polynomials,
 the calculation of an average such as (3.2) requires 
some interpolation and extrapolation of $F_2^{\rm(exp)}(x,Q^2)$. To do so 
we have used two different methods. In both we separate $F_2$ into a singlet 
and a nonsinglet part, writing
$$F_2=F_S+F_{NS}.$$
In the first method we use, 
for each 
value of $Q^2$ independently, a phenomenological 
expression for the $F$,
$$\eqalign{F_S^{\rm phen.}(x)=&(Ax^{-0.44}+C)(1-x)^\nu,\cr
F_{NS}^{\rm phen.}(x)=&B x^{0.5}(1-x)^\mu.\cr}
\equn{(3.3)}$$

\setbox3=\vbox{\hsize 12.2truecm
\setbox0=\vbox{\epsfxsize 8.truecm\epsfbox{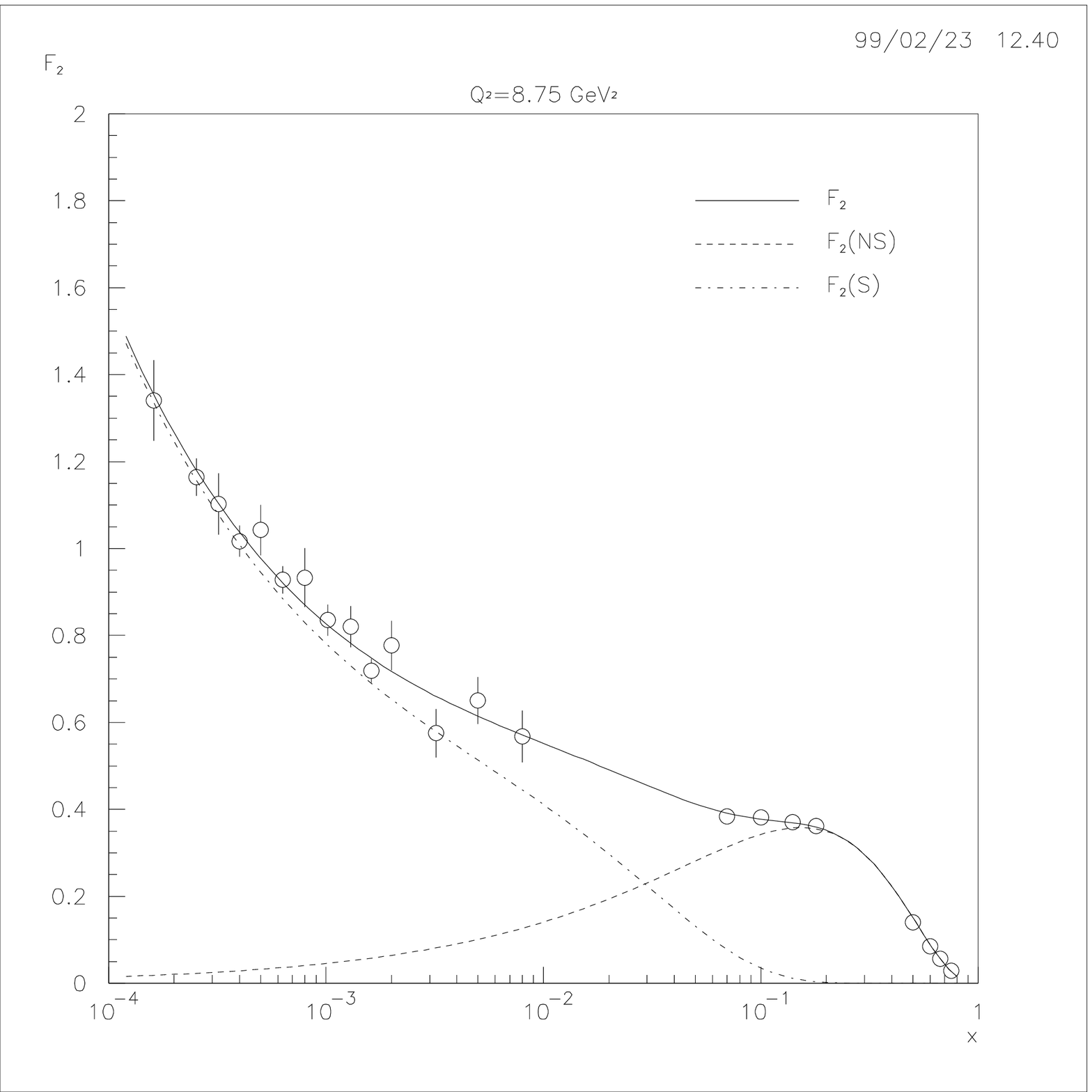}}
\setbox1=\vbox{\hsize 10.5truecm\captiontype\figurasc{Figure 1 }{Phenomenological 
fit for $Q^2=8.75\;\gev$ (logarithmic scale).}\hb
\vskip.1cm} 
\centerline{{\box0}}
\medskip
\centerline{\box1}}
\centerline{\box 3}
\setbox7=\vbox{\hsize 12.2truecm
\setbox5=\vbox{\epsfxsize 8.truecm\epsfbox{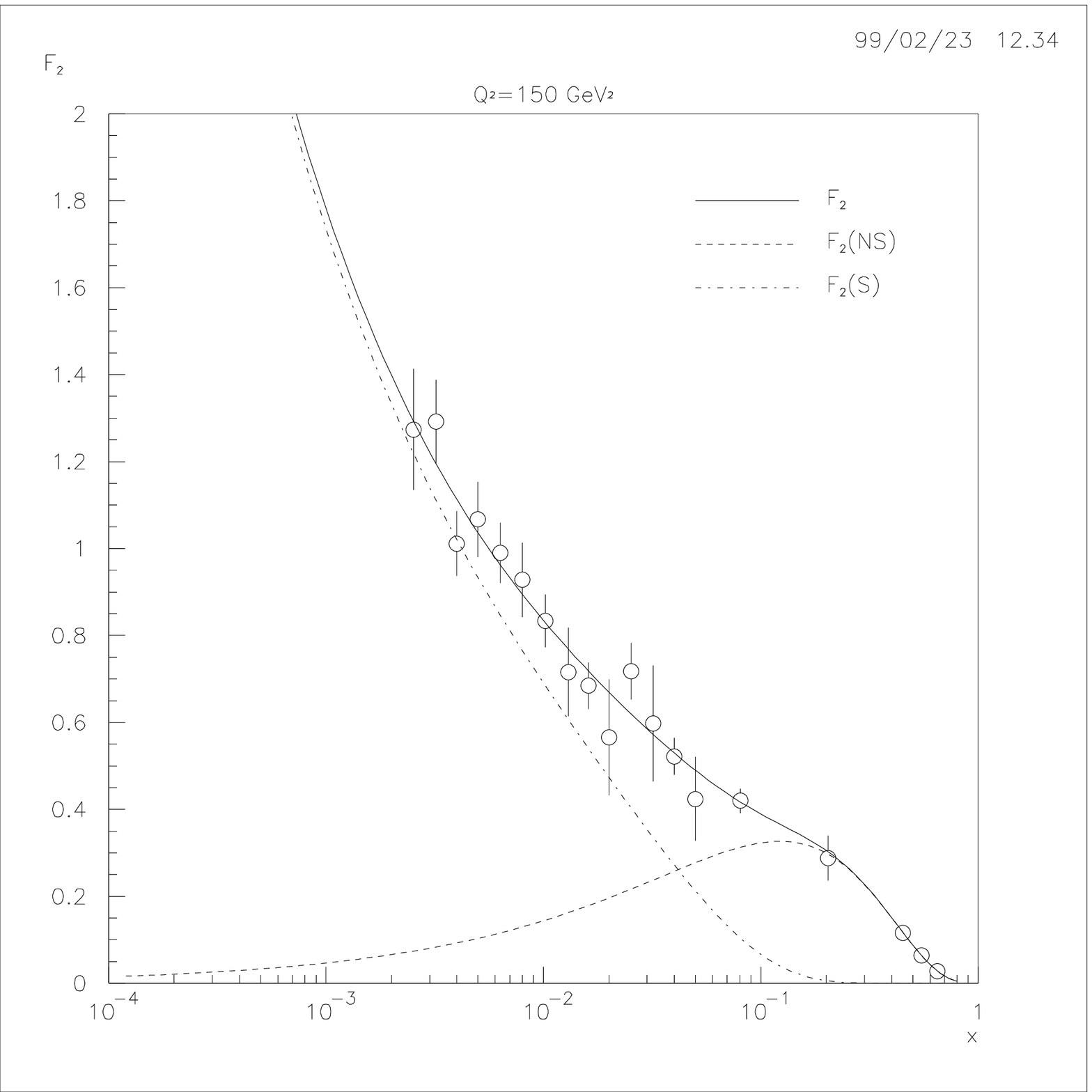}}
\setbox6=\vbox{\hsize 10.5truecm\captiontype\figurasc{Figure 2 }{Phenomenological 
fit for $Q^2=150\;\gev$ (logarithmic scale).}\hb
\vskip.1cm} 
\centerline{{\box5}}
\medskip
\centerline{\box6}}
\centerline{\box7}

A theoretical justification of (3.3) may be found in ref.~7 and 
work quoted there; but we want to emphasize that, 
in the present paper, (3.3) is to be considered as only a convenient 
interpolation of the data to allow calculation of the 
integrals (3.2). In fact, 
the parameters $A,\,B,\,C,\,\nu,\,\mu$ are taken to be 
totally free, and assumed to be {\sl uncorrelated}  
for different values of $Q^2$. Thus, no theoretical 
bias is induced in the $Q^2$ dependence.
 The fits, for a couple of representative $Q^2$ values, 
are given in \figs~1,~2. 
In \fig~3 we also show the polynomial
 $p_{31}(x)$, superimposed on the 
fit and the experimental values of $F_2(x,Q^2=8.75\,\gev^2)$,
as a representative case.

The second method we consider is to use 
the parameterization of data (that we denote by 
MRST98) given in ref.~15, which includes 
some theoretical input. Then the integral involved in the  Bernstein average 
is evaluated with the help of the parametric expression, for each value of $Q^2$. 

We consider the first method to be the cleanest one; the results 
found using the MRST98 parameterization are presented mostly 
to show the insensitivity of the 
evaluation to the method of obtaining the ``experimental" 
averages $F^{\rm(exp)}_{nk}(Q^2)$.

\setbox7=\vbox{\hsize 13.2truecm
\setbox5=\vbox{\epsfxsize 9.truecm\epsfbox{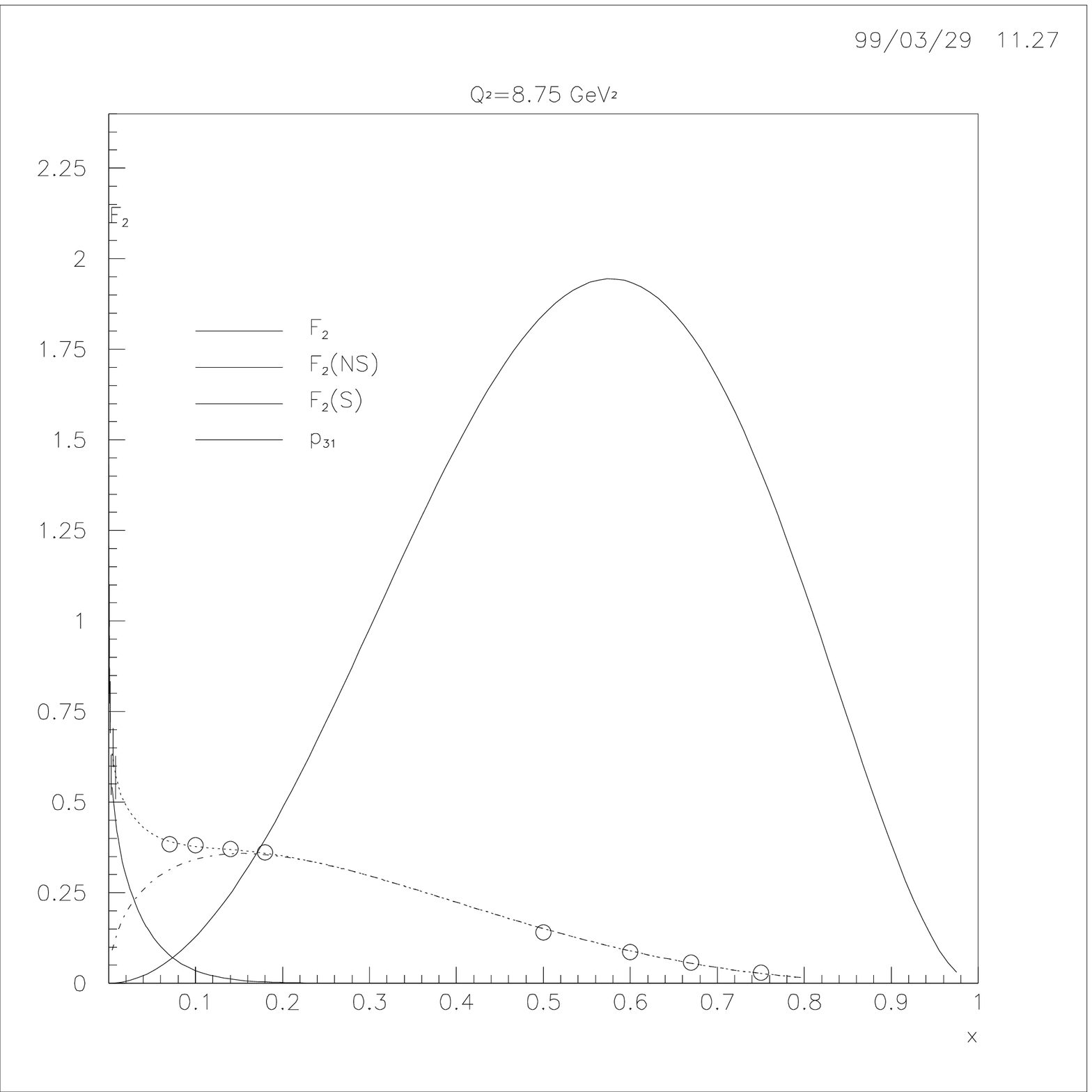}}
\setbox6=\vbox{\hsize 13.2truecm\captiontype\figurasc{Figure 3 }{Phenomenological 
fit for $Q^2=8.75\;\gev$, and Bernstein polynomial $p_{31}$.\hb Only 
this polynomial is shown as an example, although for this value of 
$Q_0^2$ other Bernstein averages are also meaningful. The points at 
very small $x$ are not shown 
(compare with \fig~2).}\hb
\vskip.1cm}
\medskip 
\centerline{{\box5}}
\medskip
\centerline{\box6}}
\medskip
\centerline{\box7}
 
\booksubsection{3.3 Experimental input}
To begin with, we have the limitation that, since the $\gamma$s in 
the QCD evolution equations are 
only known
 for $n=2,\,4,\,6,\,8$, we will be only 
able to use, for the theoretical calculations, the
 {\sl ten} Bernstein polynomials\fnote{Actually, $\gamma^{(2)}_{NS}(n)$ 
is also known for $n=10$ but we will not be able to use this because in $F_2$ for 
electroproduction singlet and nonsinglet are not separated.}
$$\eqalign{p_{nk}(x);\quad k=&0,\quad n=0,\,1,\,2,\,3;\cr
\quad k=&1,\quad n=1,\,2,\,3;\cr
\quad k=&2,\quad n=2,\,3;\cr
\quad k=&3,\quad n=3.\cr}
\equn{(3.4)}$$
However, we will not have experimental information, for a given $Q^2$, on all of 
the corresponding averages. Thus, for example, at 
very large values of $Q^2$, the HERA measurements 
have been carried over only for small values of $x$, so here 
only the  averages with $p_{30}$ and at most $p_{20}$ can 
be used. Also, there are generally 
speaking no data for $x$ near unity, so averages
 with large $n-k$ cannot be used either.
 As a matter of fact, we have only used  the  six polynomials
$$\eqalign{p_{10},\; p_{20},\; p_{30};\cr
p_{21},\;p_{31},\;p_{32}.\cr}$$

When we take this into account, it follows that we
 can calculate reliably only a total of 102 Bernstein averages, 
using the experimental data from SLAC\ref{16}, BCDMS\ref{17},
 E665\ref{18} and HERA\ref{19}. 
The experimental averages, as defined in \equn{(3.2)} with the 
interpolation given by (3.3), are shown in \fig~4 (together with 
the NNLO fit, see below); the errors there are the 
errors in the integrals (3.2) induced by the errors in 
the parameters  $A,\,B,\,C,\,\nu,\,\mu$; and these errors are
 obtained, in turn,  
from the experimental errors of the fitted $F_2$. These errors 
of $F_{nk}^{\rm (exp)}$ are what we consider the statistical, 
experimental errors of our input. There is however one more  
point that has to be discussed in this respect. We have, 
for each $Q^2$, theoretical  
information on {\sl four} moments, but we use 
{\sl six} Bernstein averages: so we are double counting part of 
our information. This is 
repaired as follows: we consider the values of $Q^2$ for which 
we can use more than four Bernstein averages, 
and consider the number above four of these to be ``duplicate" 
information. The total number 
of  these duplicates is of 34. Then we renormalize the 
experimental error by multiplying it by
 $\sqrt{102/(102-34)}= 1.22$. 

\booksection{4. Theoretical calculation. Error analysis}
The QCD  equations (2.11,~12) 
 only give the evolution of the moments. Therefore, we have to take their 
initial values, at a fixed $Q_0^2$
$$\mu_S(n,Q^2_0),\quad\mu_G(n,Q^2_0);\quad\mu_{NS}(n,Q^2_0);
 \quad n=2,\,4,\,6,\,8, $$
as  parameters.  For the actual 
calculations we chose $Q_0^2=8.75\;\gev^2$ because, for this 
value, 
there are experimental data in a wide range of values of $x$ so the 
input values are directly constrained by experiment. 
We have considered varying the value of $Q_0^2$; 
this provides an indication of the errors induced by the approximations 
made for, in an exact calculation, the result 
should be 
independent of $Q_0^2$. Indeed, the variation 
is  small; for example, if we take  $Q_0^2=12\;\gev^2$ 
the QCD mass $\Lambdav$ varies by $11\;\mev$. 
 
Of these twelve moments only eleven are free.
 In fact, because of the momentum 
sum rule it follows that 
$$\lim_{Q^2\to\infty}[\mu_S(2,Q^2)+\mu_G(2,Q^2)]=\langle Q^2_q\rangle=
\dfrac{1}{n_f}\sum_{q=1}^{n_f}Q_q^2,$$
where $\langle Q^2_q\rangle$ is the average quark charge; for $n_f=4$, 
 $\langle Q^2_q\rangle=\tfrac{5}{18}$; for  $n_f=5$, 
 $\langle Q^2_q\rangle=\tfrac{11}{45}$. To implement this, we 
may use the  
equations (2.11) with $\alpha_s(Q^2)\to0$, and that $d_+(2)=0$, $d_-(2)<0$ together 
with the values of the $S_{ij}(2)$ to get the 
constraint
$$\tfrac{5}{18}=\dfrac{16+3n_f}{3n_f}
\sum_{j=S,G}\left(\bar{M}((2,a_0)S^{-1}(2)C^{-1}(2,a_0)\right)_{1j}\mu_j(2,Q_0^2)
\equn{(4.1)}$$
and the quantities here are to 
be calculated for $n_f=4$. So we may 
use (4.1) to eliminate $\mu_G(2,Q_0^2)$ in favour of $\mu_S(2,Q_0^2)$, 
leaving  11 free moments, as stated. To these we have to add a further 
parameter viz., the QCD mass $\Lambdav$, to  a total of 12 free 
parameters for the fit.

The procedure is now rather straightforward. We calculate the $\mu(n,Q^2)$ in terms of 
the $\mu(n,Q^2_0)$ using (2.11,~12); and once these $\mu(n,Q^2)$ known,  
we evaluate the theoretical Bernstein moments $F^{\rm th}_{kn}(Q^2)$ for 
all $Q^2$, in terms of our twelve parameters.
 We are then ready to fit  the experimental quantities
 obtained before, $F^{\rm(exp)}_{nk}(Q^2)$;  
a fit which is performed with the MINUIT program\ref{20}. 
Before presenting the results, however, we have to discuss a 
number of theoretical questions, including refinements and 
 estimates of theoretical errors.

\booksubsection{4.1. Target mass corrections}
Target mass corrections (TMC) are of order $nm_p/Q^2$, for the $n$th moment ($m_p$ 
being the mass of the proton) and must 
be included in a precision calculation. 
The corrections can be evaluated exactly\ref{21}, to LO in $\alpha_s$,  
but we here will only take them into account to first order in $m_p^2/Q^2$.
 The reasons 
are that higher twist (HT) corrections, of order $n\Lambdav^2/Q^2$,
 are not known; and neither are the NLO (in $\alpha_s$) 
TMCs. Although it may be 
argued that HT are suppressed with respect to TMC 
by $1/N_c$, with $N_c=3$ 
the number of colours in the large $N_c$ 
limit, it is useless  
to include effects $O(m_p^4/Q^4)$ 
while the ones of order $\Lambdav^2/Q^2$ and $O(\alpha_s m_p^2/Q^2)$ 
corrections  are not known. Anyway, we will give results both 
with and without TMC; the variation is slight, and the 
error, as just estimated, negligible.

We write (see  refs.~12,~21) 
$$\mu^{TMC}_{NS}(n,Q^2)=\mu_{NS}(n,Q^2)+\dfrac{n(n-1)}{n+2}
\dfrac{m_p}{Q^2}\mu_{NS}(n+2,Q^2).
\equn{(4.2)}$$
It should be noted that the TMCs for the singlet are negligible 
compared to those of the NS, which is why we only take 
into account the last. Another point is 
that (4.2) involves one further moment, $\mu_{NS}(10,Q^2)$.
 Since we do not want to add 
a new parameter to the fit, what we do is to {\sl estimate} 
 $\mu_{NS}(10,Q^2)$ from the fitted expression for $F_2$, 
\equn{(3.3)}. Since this is only a correction, 
it does not matter much if it is not very accurate.

\booksubsection{4.2. Correlation between gluon structure 
function and $\Lambdav$}
As is known, in other fits there is a strong 
correlation between the value of the QCD parameter $\Lambdav$ 
and the gluon structure function, which leads to 
instabilities in the values of the last; this is due to the 
fact that the $\mu_G$ are not 
directly related 
to a measured quantity. The problem is however not 
serious in our case. This is because, in 
our evaluations, $xG(x,Q^2)$ is only 
 represented  by its 
three input moments
$$\mu_G(4,Q_0),\quad\mu_G(6,Q_0),\quad\mu_G(8,Q_0);
$$
recall that $\mu_G(2,Q_0)$ can be eliminated in favour of $\mu_S(2,Q_0)$ using 
the momentum sum rule.  
We have also found that the instabilities disappear completely if 
we simply make the requirement that $\mu_G(n,Q^2)>\mu_G(n+2,Q^2)$, 
a requirement that follows from the positivity of $G(x,Q^2)$ as a 
probability density for gluons with 
momentum fraction $x$. In fact, from the positivity of $G$ we can 
impose other inequalities;\ref{14} for example, considering only linear ones, 
we must have
$$\eqalign{\int_0^1d x\,p_{nk}(x)F_G(x,Q_0^2)=&
\dfrac{2(n-k)!\Gamma(n+\tfrac{3}{2})}{\Gamma(k+\tfrac{1}{2})\Gamma(n-k+1)}
\sum_{l=0}^{n-k}\dfrac{(-1)^l}{l!(n-k-l)!}\,\mu_{G}(2k+2l+2,Q_0^2)\geq 0.\cr}$$
We have checked that his is verified by the central 
values for the $\mu_G$ 
in our fits. 
 
If we allowed the $\mu_G$ to vary freely 
this would result in the appearance of a spurious minimum 
for which the lowest moment would be smaller 
than the 
higher ones. Nevertheless, even in this case 
the value of $\Lambdav$ does not 
vary too much: it only decreases by some $10\;\mev$.  
In our fits we of course require  positivity.

\booksubsection{4.3. Quark mass dependence}
Since our $Q^2$ range goes through the 
$b$ quark mass threshold, we have to worry about that region. In our 
main calculation we have simply used the matching conditions (2.13), 
fixing the mass of the $b$ quark from 
the recent determination in ref.~22, correct to $O(\alpha_s^4)$,
$$m_b=5\,001^{+101}_{-66}\;\mev.$$
It is to be noted, however, that the use of 
(2.13) is not enough to take into account exactly 
the mass dependence. To do so, one would have to rewrite (and solve) the 
renormalization group equations taking into account the 
finiteness of $m_b$, a very hard task 
quite outside our scope here. The fact 
that the use of (2.13) gives only an approximation to the mass dependence 
results in a discontinuity of the theoretical calculation, at 
NNLO, near $Q^2=m_b^2$, which can be seen 
clearly in Fig.~4. Since the discontinuity 
is of  $\lsim1\%$, and affects only 
the fit to very few 
experimental points, we consider the use of (2.13) to 
be the best available procedure. 
 Alternatively, 
we could simply avoid the region around the $b$ quark threshold and thus
 cut off the interval\fnote{We could also consider increasing the minimum 
value of $Q^2$ to be well clear of 
the $c$ quark threshold, but we have not bothered to do so 
as the experimental points in the corresponding region are not significant; cf. \fig~4.} 
$$m_b^2\leq Q^2\leq 4m_b^2.$$
Actually, we remove from the fit the values of the moments for $Q^2$ 
between $20$ and $90\;\gev^2$.
Again, we take the difference between the results  with both methods
 (matching or cutting the threshold) as part of the 
measure of the error due to the $b$ quark mass effect. The 
rest of the error is obtained varying $m_b$ inside its error bars, as above.

\booksubsection{4.4. Higher order corrections}
Higher order corrections are of two types. First, we have higher order in 
$\Lambdav^2/Q^2$, viz., higher twist effects (HT). Since so little 
is known about these we simply take into account 
phenomenologically the more important ones by adding, to
$\mu_{NS}(n,Q^2)$, the  
correction  
$$\mu_{NS}^{HT}(n,Q^2)=n(a\Lambdav^2/Q^2)\mu_{NS}(n,Q^2)
\equn{(4.3)}$$ 
with $a$ an unknown parameter, to be fitted, and expected 
to be of order unity; in 
our fit we find $a\sim -0.20$, see below. The difference between the results with this new 
contribution, and the one without, will be the 
estimated error induced by HT effects. Note that 
our central values (Table 1 below) are 
obtained {\sl without} including the HT 
term. This is because, although 
the value we get for $a$ is reasonable, the fact 
remains that the expression (4.3) used for HT 
is little more than educated guesswork, and we  want to 
avoid as much as possible to introduce biases in our calculation 
 of $\Lambdav$, $\alpha_s$. 

The second type of higher order 
corrections will be corrections of relative 
order $\alpha_s^3$, i.e., {\sl N}\/NNLO corrections. We estimate these 
by the different results obtained 
performing our calculations either with the fractional expression (2.12b), or with 
the expanded denominator one, (2.12c). This simulates, at least in 
the expected behaviour at large $n$, the largest NNNLO 
corrections; in fact, the largest corrections of 
order $j$  to $F_2$ may be argued to behave as $\alpha_s^j\log^j (1-x)$ and 
stem from the  $C_{NS}^{(j)}(n)$, $\gamma_{NS}^{(j)}(n)$ at large $n$; see 
for example ref.~12, \sect~4.9ii and references therein. By 
using our procedure, we are estimating e.g.,
$$C_{NS}^{(3)}(n)\sim C_{NS}^{(2)}(n)C_{NS}^{(1)}(n).$$

\booksubsection{4.5. Dependence on parameterizations}
The dependence of our input data for the Bernstein averages 
on the parameterizations used to evaluate the 
corresponding integrals is estimated as follows: we calculate using 
our interpolation formulas, \equn{(3.3)}, 
and then repeat the calculation with the MRST98 parameterization. 
The difference (10 \mev\ for $\lambdav$) will be our estimated error 
due to this source.

\setbox7=\vbox{\hsize 12.9truecm
\setbox5=\vbox{\epsfxsize 8.9truecm\epsfbox{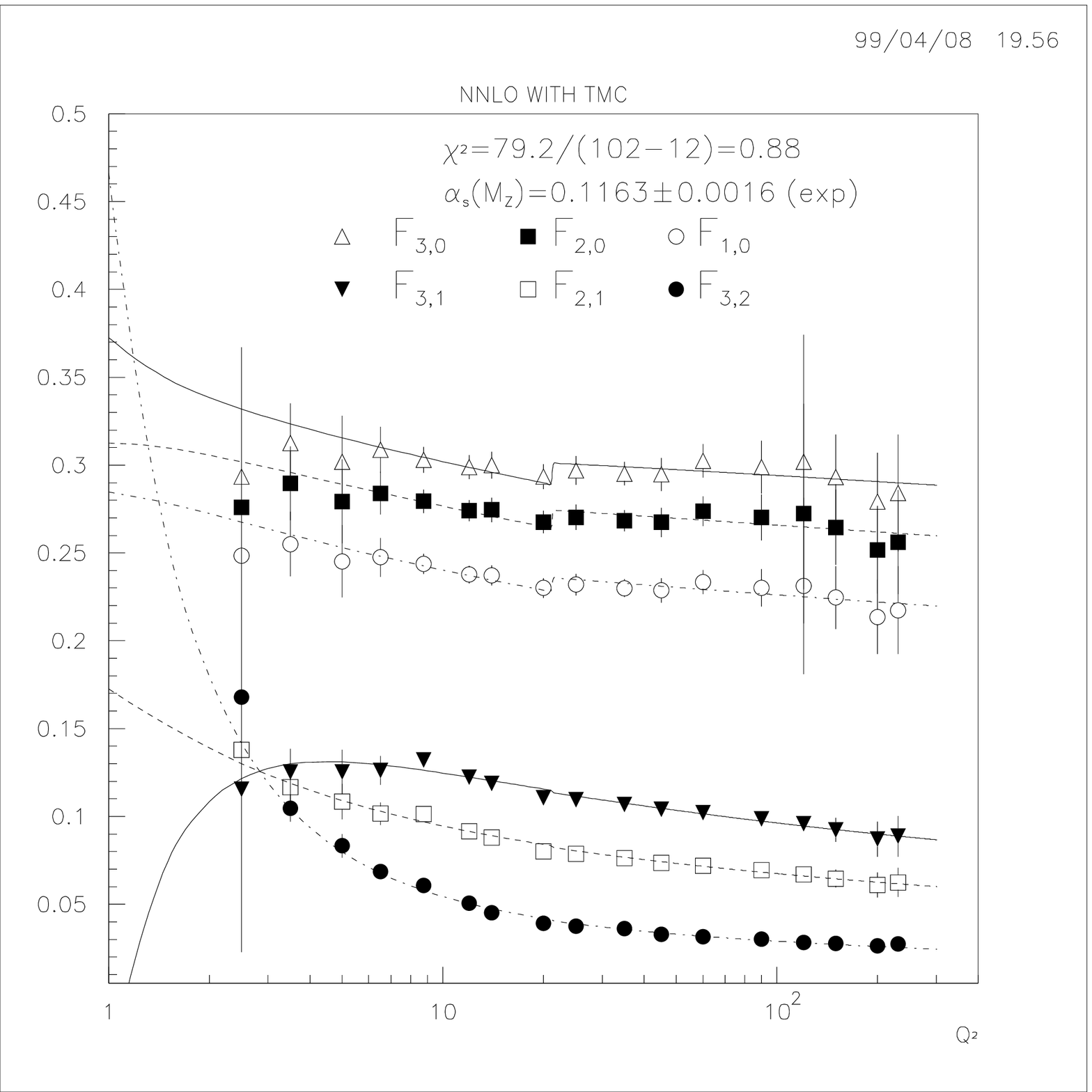}}
\setbox6=\vbox{\hsize 12.8truecm\captiontype\figurasc{Figure 4 }{Experimental 
values for the Bernstein averages, and NNLO 
theoretical curves, with 
the gluon structure function constrained and TMC 
taken into account.}\hb
\vskip.1cm} 
\centerline{{\box5}}
\medskip
\centerline{\box6}}
\medskip
\centerline{\box7}
\medskip

\booksection{5. Results. Discussion}
We  present in Table 1 a compilation of the results obtained
 with our calculations at LO, NLO and 
NNLO, with TMCs taken into account; the fit to the data itself is shown,
 for the NNLO calculation (with TMC) in \fig~4. In this Table 1, 
only {\sl statistical} errors are shown; systematic (theoretical) errors will 
be discussed below. We note that 
the errors given in Table 1 are ``renormalized" 
to take into account the effective number of independent 
experimental points, 
as discussed at the end of \sect~3. That is to say, 
the error presented is obtained from the raw error by multiplying it by 1.22. 
In calculating the \chidof, however, we have considered all
 the input experimental points, as both the total chi-squared and 
the number of degrees of freedom change by 
the same amount. 

\setbox0=\vbox{
\setbox1=\vbox{\offinterlineskip\hrule
\halign{
&\vrule#&\strut\hfil#\hfil&\vrule#&\strut\hfil#\hfil&
\quad\vrule\quad#&\strut\quad#\quad&\quad\vrule#&\strut\quad#\cr
 height2mm&\omit&&\omit&&\omit&&\omit&\cr 
&\phantom{l}Order\phantom{l}&&\phantom{l}$\Lambdav(n_f=4)$&
&$\alpha_s(M_Z^2)$\kern.3em&&\chidof& \cr
 height1mm&\omit&&\omit&&\omit&&\omit&\cr
\noalign{\hrule} 
height1mm&\omit&&\omit&&\omit&&\omit&\cr
&LO&&\phantom{\big|}$ 215\pm73$&&$0.130\pm0.005$&&$212/(102-12)\;$& \cr
\noalign{\hrule} 
height1mm&\omit&&\omit&&\omit&&\omit&\cr
&NLO&&\phantom{\big|}$ 282\pm40$&&$0.116\pm0.0034$&&$80.0/(102-12)\;$& \cr
\noalign{\hrule} 
height1mm&\omit&&\omit&&\omit&&\omit&\cr
&NNLO&&\phantom{\big|}$ 283\pm25$&&$0.1172\pm0.0017$&&$79.2/(102-12)\;$& \cr
 height1mm&\omit&&\omit&&\omit&&\omit&\cr
\noalign{\hrule}}
\vskip.05cm}
\centerline{\box1}
\smallskip
\centerline{\petit Table 1}
\centerrule{6cm}
\medskip}
\box0
The NLO corrections are very clearly seen in the 
fit: the 
\chidof\ decreases from a largish value of $\sim2.4$ to a very good $\sim0.89$ 
when including these. The fit is so good already at 
this order that there is very little room for improvement when going 
to NNLO; nevertheless, an improvement is seen. Not 
in the \chidof, which only decreases 
minutely (to  $\chidof \sim0.88$) but certainly in that
 including NNLO corrections leads to 
 a noticeable gain both in the 
quality of the determination of the coupling, and in the stability of the fits.

The estimated systematic errors, originating  from various sources,
 are  shown for the NNLO case  
in Table 2:

\setbox0=\vbox{
\medskip
\setbox1=\vbox{\offinterlineskip\hrule
\halign{
&\vrule#&\strut\hfil#\hfil&\vrule#&\strut\hfil#\hfil&
\quad\vrule\quad#&\strut\quad#\quad&\quad\vrule#&\strut\quad#\cr
 height2mm&\omit&&\omit&&\omit&&\omit&\cr 
&\phantom{l}Source of error\phantom{l}&&\phantom{l}$\Lambdav(n_f=4;\hbox{3 loop})$&
&\phantom{l}$\lap\Lambdav(n_f=4;\hbox{3 loop})$&&$\lap\alpha_s(M_Z^2)$\kern.3em& \cr
 height1mm&\omit&&\omit&&\omit&&\omit&\cr
\noalign{\hrule} 
height1mm&\omit&&\omit&&\omit&&\omit&\cr
&No TMC&&\phantom{\big|}$292$&&\hfil$9$\hfil&&$0.0006$& \cr
\noalign{\hrule} 
height1mm&\omit&&\omit&&\omit&&\omit&\cr
&\phantom{\big|}Interpol. 
(MRST98)\phantom{\big|}&&\phantom{\big|}$ 273$&&\hfil$10$\hfil&&$0.0007$& \cr
\noalign{\hrule} 
height1mm&\omit&&\omit&&\omit&&\omit&\cr
&HT&&\phantom{\big|}$292$&&\hfil$9$\hfil&&$0.0006$& \cr
\noalign{\hrule} 
height1mm&\omit&&\omit&&\omit&&\omit&\cr
&Quark mass effect&&299&&\hfil$16$\hfil&&$0.0011$& \cr
\noalign{\hrule}
height1mm&\omit&&\omit&&\omit&&\omit&\cr
&$Q_0^2$ to $12\;\gev^2$&&\phantom{\big|}$294$&&\hfil$11$\hfil&&$0.0008$& \cr
\noalign{\hrule}
height1mm&\omit&&\omit&&\omit&&\omit&\cr
&{\sl N}NNLO&&\phantom{\big|}$289$&&\hfil$6$\hfil&&$0.0004$& \cr
 height1mm&\omit&&\omit&&\omit&&\omit&\cr
\noalign{\hrule}}
\vskip.05cm}
\centerline{\box1}
\smallskip
\centerline{\petit Table 2}
\centerrule{6cm}
\medskip}
\box0
\noindent Let us comment on the meaning of the different entries. 
No TMC means that we have {\sl not} taken 
target mass corrections into account. The 
corresponding  
error is {\sl not} included when evaluating the overall theoretical error 
because, since we take into account TMC in 
our central value, the error 
would be of order $TMC^2$, or $\alpha_s\times TMC$, 
quite negligible. 
 MRST98 means that we have used the MRST98 interpolation (\subsect~3.2)  
to calculate the integrals (3.2).  
 HT means that we have 
taken into account the presence of higher 
twist by adding 
a contribution like  (4.2). The fitted value of the 
phenomenological parameter $a$ is $a=-0.202\pm0.030$.\fnote{This 
is a very reasonable value that makes the mass associated with 
higher twists, $M_{HT}=\Lambdav\sqrt{|a|}\sim0.13\,\gev$ of 
the order of typical HT parameters: intrinsic average 
transverse momentum of the quarks, or inverse proton 
radius; see e.g. 
the paper of De R\'ujula et al., ref.~1.} 
 ``Quark mass effect" means that we have 
cut off the $b$ quark threshold as discussed in \subsect~4.3; 
the error in Table 2 takes into account 
also the variations due to the error in the 
$m_b$ mass. $Q_0^2$ to $12\,\gev^2$ means 
that we take the input moments defined at this 
value of the momentum, $\mu_i(n,Q_0^2=12\,\gev^2)$, $i=S,\,G,\,NS$.
 Finally, {\sl N}NNLO means that we have fitted with 
the expanded formula (2.12c) instead of the 
fractional one (2.12b). Note that we only 
take this into account for the nonsinglet. This is because for large $n$, which is 
when the higher order corrections are larger, the NS 
piece of the structure function is the dominating one.

Composing quadratically systematic (theoretical) and statistical (experimental)  
errors we find the best result for the QCD coupling, 
$$\eqalign{\Lambdav(n_f=4,\,\hbox{3 loop})=&282.7\pm25.1\;(\hbox{stat.})\pm24.5\;(\hbox{syst.})
=282.7\pm35.1\;\mev;\cr 
 \alpha_s^{(\rm 3\, loop)}(M_Z)=&0.1172\pm0.0017\;(\hbox{stat.})\pm0.0017\;(\hbox{syst.})
=0.1172\pm0.0024;\cr} 
\equn{(5.1)}$$
the corresponding central value for $\lambdav(n_f=5,\,\hbox{3 loop})$ 
is of $0.200\,\mev$. It is to be noted that composing the ``theoretical" 
errors 
as if they were independent leads to a certain amount of double-counting. Thus, the 
results should be independent of the value 
of $Q_0^2$ if the calculation was to all orders, so the 
two last errors in Table 2 are connected. 
We have, however, preferred to play it safe, 
particularly because these errors are in some cases 
little more than rough estimates.

We next show a table comparing our results to previous determinations
 for $\alpha_s(M_Z^2)$; these are taken from the review by Bethke\ref{23},
including only the processes where the theoretical calculation
 has been pushed to the NNLO level 
and the experimental data are good enough to make the
 analysis meaningful.\fnote{This excludes $e^+e^-$ annihilations.} 
We also incorporate in the table  
 the recent results in ref.~10:

\setbox0=\vbox{
\medskip
\setbox1=\vbox{\offinterlineskip\hrule
\halign{
&\vrule#&\strut\hfil#\hfil&\quad\vrule\quad#&
\strut\quad#\quad&\quad\vrule#&\strut\quad#\cr
 height2mm&\omit&&\omit&&\omit&\cr 
& \kern.5em Process&&${\textstyle\hbox{Average}\;
 Q^2}\atop{\textstyle \hbox{or}\; Q^2\;\hbox{range}\;[\gev]^2}$&
& $\alpha_s(M_Z^2)$\kern.3em& \cr
 height1mm&\omit&&\omit&&\omit&\cr
\noalign{\hrule} 
height1mm&\omit&&\omit&&\omit&\cr
&\phantom{\Big|}  DIS; $\nu$, Bj&&2.5&&0.122& \cr
\noalign{\hrule}
&\phantom{\Big|}  DIS; $\nu$, GLS&&3&&0.115& \cr
\noalign{\hrule}
&\phantom{\Big|}  $\tau$ decays&&$(1.777)^2$&&0.119&\cr
\noalign{\hrule}
&\phantom{\Big|}  $Z\to{\rm hadrons}$&&$(91.2)^2$&&0.124&\cr
\noalign{\hrule}
&\phantom{\Big|}  DIS; $\nu,\;xF_3$&&$5 - 100$&&$0.117\pm0.010$\phantom{l}& \cr
\noalign{\hrule}
&\phantom{\Big|}  our result&&${2.5} - {230}$&&$0.1172\pm0.0024$\phantom{l}&\cr
 height1mm&\omit&&\omit&&\omit&\cr
\noalign{\hrule}}
\vskip.05cm}
\centerline{\box1}
\smallskip
\centerline{\petit Table 3}
\centerrule{6cm}
\medskip}
\box0
\noindent Here DIS means deep inelastic scattering, 
Bj stands for the Bjorken, and GLS for the Gross--Llewellyn Smith 
sum rules. The $xF_3$ result is that of ref.~10.

The previously existing average value,
 also taking into account NLO calculations, was
$$\alpha_s(M_Z^2)=0.118\pm0.006;$$
when including both our result and that of ref.~10 the new average
and error become
$$\alpha_s(M_Z^2)=0.1174\pm0.0016.\equn{(5.2)}$$

We add   a further comment. 
If we only kept, in the old determinations 
of $\alpha_s$, processes with {\sl spacelike} momenta
 then, as noted by Bethke,\ref{23} 
a slightly smaller value was obtained for the coupling:
$$\alpha_s(M^2_Z)=0.114\pm0.005\;
(\hbox{spacelike momenta}).$$
Our results correct this to some extent, 
by increasing slightly the average value for ``spacelike" determinations. 

To finish this paper, we discuss briefly why our calculation 
yields such accurate results. First of all, 
and compared with neutrino data, our experimental input 
turns out to be much more 
precise, and also it extends over a much wider range of $Q^2$ 
values. Actually, this wide range of 
values is one of 
the assets in our evaluation, particularly when comparing it to 
$Z,\,\tau$ decays where only one value of $Q^2$ is essentially 
available. Our range is such that not 
only leading logs, but next-to-leading logs 
vary appreciably: cf Table 1. The inclusion 
of NNL logs then stabilizes the results and decreases 
the errors and  
(very slightly) the \chidof.

\vfill\eject
\booksection{Acknowledgments}
The authors are indebted to A. Vermaseren for
 several illuminating discussions, and to L. Labarga for the same reason and for 
informatic help. Prof. Bl\"umlein has also contributed 
with useful remarks about target mass corrections. 
Finally, thanks are also due to S.~Bethke who spotted (and 
allowed us to correct)  
a mistake in the paper.

 The financial support of CICYT (Spain, contracts 
AEN 97-1678 and AEN 96-1672) is  gratefully acknowledged.
 One of us (J. S.) wishes also to thank the Junta de Andaluc\'\i a, MEC 
and IEM. 

\booksection{References}
\item{1.- }{\ajnyp{D. J. Gross and F. Wilczek}{Phys. Rev.}{D9}{1974}{980}; 
\ajnyp{H. Georgi and H. D. Politzer}{Phys. Rev.}{D9}{1974}{416}; 
\ajnyp{I. Hinchliffe and 
C. H. Llewellyn Smith}{Nucl. Phys.}{B128}{1977}{93};
 \ajnyp{A.  De R\'ujula, H. Georgi
 and H. D. Politzer}{Ann. Phys. ({\rm NY})}{103}{1977}{315}.}
\item{2.- }{\ajnyp{A. Gonz\'alez-Arroyo,  C. L\'opez and
 F. J. Yndur\'ain}{Nucl.
 Phys.}{B153}{1979}{161}; 
{\bf B159} (1979) 512;{\bf B174},
 (1980) 474.}
\item{3.- }{\ajnyp{E. G. Floratos, D. A. Ross and C. T.
 Sachrajda}{Nucl. Phys.}{B129}{1978}{66}; 
(E) {\bf B139} (1978) 545.}
\item{4.- }{\ajnyp{W. A. Bardeen et al.}{Phys. Rev.}{D18}{1978}{3998}.}
\item{5.- }{\ajnyp{D. J. Gross}{Phys. Rev. Lett.}{32}{1974}{1071}.}
\item{6.- }{\ajnyp{W. Furmanski and R. Petronzio}{Phys. Lett.}{97B}{1980}{437}.}
\item{7.- }{\ajnyp{K. Adel, F. Barreiro and F. J. Yndur\'ain}
{Nucl. Phys.}{B495}{1997}{221}.}
\item{8.- }{\ajnyp{W. L. van Neerven and E. B. Zijlstra}
{Phys. Lett.}{B272}{1991}{127 and 476}; 
{\sl ibid} { B273} {(1991)} {476}; {\sl Nucl. Phys.} {\bf B383} (1992) 525.}
\item{9.- }{\ajnyp{S. A. Larin et al.}{Nucl. Phys.}{B427}{1994}{41} and 
{\bf B492} (1997), 338.}
\item{10.- }{\kern0.25em{\sc A. L. Kataev, G. Parente and A. V. Sidorov},
 JINR E2-98-265 (1998).}
\item{11.- }{\ajnyp{F. J. Yndur\'ain}{Phys. Lett.}{74B}{1978}{68}.}
\item{12.- }{\kern0.25em{\sc F. J. Yndur\'ain}, 
{\sl The Theory of Quark and Gluon Interactions}, 
3rd. Ed.,  Springer--Verlag, New York, 1999.}
\item{13.- }{\ajnyp{S. A. Larin, 
T. van Ritbergen and J. A. M. Vermaseren}{B438}{1995}{278}; 
\ajnyp{K. G. Chetyrkin et al.}{Phys. Rev. Lett.}{79}{1997}{2184}.}
\item{14.- }{\kern0.25em{\sc F. J. Yndur\'ain}, 
{\sl The Moment Problem and Applications}, Int'l School on Pad\'e Approximants,
 The British Institute of Physics, 1973.}
\item{15.- }{\ajnyp{A. D. Martin et al.}{Eur. Phys. J.}{C4}{1998}{463}.}
\item{16.- }{\ajnyp{L. W. Whitlow et al.}{Phys. Lett.}{B282}{1992}{475}.}
\item{17.- }{\ajnyp{A. Benvenuti et al.}{Phys. Lett.}{B223}{1989}{485}.}
\item{18.- }{\ajnyp{M. R. Adams et al.}{Phys. Rev.}{D54}{1996}{3006}.}
\item{19.- }{\ajnyp{M. Derrick et al.}{Z. Phys.}{C72}{1996}{399} and 
\ajnyp{S. Aid et al.}{Nucl. Phys.}{B470}{1996}{3}.}
\item{20.- }{\kern0.25em{\sc F. James}, {\sl CERN Program Library
 Long Writeup D506}, version 94.1 (1994).}
\item{21.- }{\ajnyp{O. Nachtmann}{Nucl. Phys.}{B63}{1973}{237}; 
\ajnyp{H. Georgi and H. D. Politzer}{Phys. Rev.}{D14}{1976}{1829}.}
\item{22.- }{\ajnyp{A. Pineda and F. J. Yndur\'ain}{Phys. Rev.}{D58}{1998}{094022} 
and CERN-TH 98-402 (hep-ph/9812371).}
\item{23.- }{\ajnyp{S. Bethke}{Nucl. Phys. Proc. Suppl.}{B64}{1998}{54}.}

\bookendchapter
\bye